\documentclass[prb,preprint]{revtex4-1} 

\usepackage{amsmath}  
\usepackage{amsfonts} 
\usepackage{graphicx} 

\newcommand{\ba}{\begin{eqnarray}}
\newcommand{\ea}{\end{eqnarray}}
\newcommand{\bvec}[1]{\mbox{\boldmath $#1$}}
\newcommand{\ket}[1]{\vert #1 \rangle}
\newcommand{\bra}[1]{\langle #1 \vert}

\begin{document}


\title{
WKB and ``cubic-WKB'' methods as an adiabatic approximation
}

\author{Shinji Iida}
\email{iida@rins.ryukoku.ac.jp} 
\affiliation{Department of Applied Mathematics and Informatics,
\\
Ryukoku University, Otsu, Shiga 520-2194, 
Japan 
}



\date{\today}

\begin{abstract}
This paper shows that WKB wave function can be expressed 
in the form of an adiabatic expansion. 
To build a bridge between two widely invoked approximation schemes
seems pedagogically instructive. 
Further, ``cubic-WKB'' method that has been devised in order to overcome
the divergence problem of WKB can be also presented in the form of 
an adiabatic approximation: 
The adiabatic expansion of a wave function contains a certain parameter. 
When this parameter is adjusted so as to make the next order 
correction vanish approximately, 
the adiabatic wave function becomes equivalent to that of the ``cubic-WKB''.
\end{abstract}

\maketitle 

\section{Introduction} 

WKB method has been widely used in various fields of physics and chemistry.
\cite{BerryMount, Miller}
It has also been a reservoir of new ideas, 
e.g. the supersymmetric WKB,\cite{SupersymmetryWKB,SupersymmetryWKB2}   
the exact WKB analysis \cite{ExactWKB,Resurgence} and so on.
In this paper we show that WKB method 
can be formulated as a special case of the adiabatic approximation, 
another widely invoked approximation scheme.
To build a bridge between two seemingly different methods
is pedagogically instructive. 
It may also be useful to 
improve both methods 
because findings in the one side could be brought to 
the other side.
Further, the ``cubic-WKB'' method \cite{cubicWKBmethod,cubicWKBtheory} 
that has been devised in order to 
overcome the divergence problem of WKB is also able to be presented in the form of an adiabatic approximation:
The adiabatic expansion of a wave function contains a certain parameter.
When this parameter is adjusted so as to make the next order 
correction vanish approximately,
the adiabatic wave function becomes equivalent to that of 
the ``cubic-WKB'' method.

\section{Adiabatic approximation}

Let us start with a brief summary of the adiabatic approximation for later use.
\cite{Messiah, GeometricPhase}
Consider the following linear differential equation:
\begin{equation}
 \frac{d\bvec{y}(x)}{dx} = M(x) \bvec{y}(x) 
\,,
\label{dydx}
\end{equation}
where $\bvec{y}$ is an $N$-dimensional vector and 
$M$ is an $N\times N$ matrix that depends on $x$ but not on $\bvec{y}$.
Adiabatic eigenvalues and eigenvectors are given as follows:
\begin{equation}
 M(x) \ket{n(x)} = \lambda_n(x) \ket{n(x)}
\,.
\end{equation}
The expansion of $\bvec{y}$ in terms of $\{ \ket{n}\}$,
\begin{equation}
  \bvec{y}(x) = \sum_{n=1}^{N} c_{n}(x) \ket{n(x)}
\,,
\label{ExpandionOfy}
\end{equation}
is inserted into Eq.(\ref{dydx}), and
we get the equations for $\{ c_n(x) \}$:
\begin{equation}
 \frac{d c_n(x)}{dx} = \lambda_{n}(x) c_{n}(x) 
- \sum_{\ell=1}^{N} \bra{n(x)}\frac{d}{dx}\ket{\ell(x)} c_{\ell}(x)
\,,
\label{dcndx}
\end{equation}
where $\{\bra{n(x)}\}$ are dual eigenvectors:
\begin{equation}
 \bra{m(x)} M(x) = \lambda_m(x)\bra{m(x)}\,,
\end{equation}
\begin{equation}
 \bra{m(x)}n(x) \rangle = \delta_{m,n}
\,.
\end{equation}
Expecting slow variations of $M(x)$, we introduce 
a small parameter $\epsilon$ by changing $x$ to $\tau$ as
\begin{equation}
 \tau = \epsilon x
\,.
\end{equation}
Then, Eq.(\ref{dcndx}) becomes
\begin{equation}
 \epsilon \frac{d c_n}{d\tau} = \lambda_{n} c_{n} 
- \epsilon \sum_{\ell=1}^{N} \bra{n}\frac{d}{d\tau}\ket{\ell} c_{\ell}
\,.
\label{dcndtau}
\end{equation}
We imagine $c_1$ is dominant and all the 
other coefficients $\{c_{\ell}; \, \ell=2,\cdots ,N\}$ are small:
\begin{equation}
c_1 \sim {\cal{O}}(1)\,,\quad 
c_{\ell} \sim {\cal{O}}(\epsilon) 
\,,\quad \ell=2,\cdots ,N
\,,
\end{equation}
and assume following expansions:
\ba
 c_1 &=& \exp
\left( \frac{1}{\epsilon} S_{0} + S_{1} + \epsilon S_2 + \cdots \right)
\,,
\label{ExpansionOfc1}
\\
c_{\ell} &=& \epsilon \Big( f_{1}(\ell) + \epsilon f_{2}(\ell) + \cdots 
\Big) c_1\,,\quad \ell=2,\cdots ,N
\,.
\label{ExpansionOfcell}
\ea
Inserting Eqs. (\ref{ExpansionOfc1}) and (\ref{ExpansionOfcell}) 
into Eq.(\ref{dcndtau}), and equating both sides of Eq.(\ref{dcndtau}) in 
each order of $\epsilon$, 
we can sequentially get following expressions:
\ba
S_{0} &=& \int^{\tau} \lambda_1 d\sigma
= \epsilon\int^{x} \lambda_{1} ds
\,,
\label{S0}
\\
S_{1} &=& - \int^{\tau} \bra{1}\frac{d}{d\sigma}\ket{1} d\sigma
= - \int^{x} \bra{1}\frac{d}{ds}\ket{1}ds
\,,
\label{S1}
\\
S_{2} &=& - \sum_{m=2}^{N}\int^{\tau} 
\frac{\bra{1}\frac{d}{d\sigma}\ket{m}\bra{m}\frac{d}{d\sigma}\ket{1}}
{\lambda_{m} - \lambda_{1}} d\sigma
\nonumber\\
&=& - \frac{1}{\epsilon}
\sum_{m=2}^{N}\int^{x} 
\frac{\bra{1}\frac{d}{ds}\ket{m}\bra{m}\frac{d}{ds}\ket{1}}
{\lambda_{m} - \lambda_1} ds
\,,
\label{S2}
\ea
where $\sigma = \epsilon s$, and
\begin{equation}
  f_1(\ell) = \frac{\bra{\ell}\frac{d}{d\tau}\ket{1}}{\lambda_\ell - \lambda_1} = \frac{1}{\epsilon}
\frac{\bra{\ell}\frac{d}{dx}\ket{1}}{\lambda_\ell - \lambda_1}
\,,\quad \ell = 2, \cdots , N
\,.
\label{f1}
\end{equation}
When Eq.(\ref{dydx}) is a time-dependent Schr\"{o}dinger equation, 
$S_0$ gives the adiabatic dynamical phase and $S_1$ is the origin of 
so-called Berry's phase.\cite{GeometricPhase,Holstein}

\section{WKB and ``cubic-WKB'' methods}

We will first show that a
usual WKB wave function is obtained in the form of 
Eq.(\ref{ExpansionOfc1}).
Consider 
one-dimensional time-independent Schr\"{o}dinger equaion
for a particle with a mass $m$ and an energy $E$ under
a potential $V(x)$:
\begin{equation}
  \psi(x)'' = - k^2(x) \psi(x)\,,\quad 
k(x) = \sqrt{\frac{2m}{\hbar^2 }( E - V(x))}
\,,
\label{SchroedingerEq}
\end{equation}
where and hereafter the prime means differentiation with respect to $x$.
Let us introduce following functions:
\begin{equation}
  y_0(x) = \psi(x)\,,\quad y_1(x) = \frac{d\psi(x)}{dx}
\,.
\end{equation}
Then, Eq.(\ref{SchroedingerEq}) becomes
\begin{equation}
  \frac{d}{dx}
\left(
\begin{array}{c} y_0(x)
\\
 y_1(x) 
\end{array}
\right)
= \left(
\begin{array}{cc}
0 & 1 \\
- k^2(x) & 0
\end{array}\right)
\left(
\begin{array}{c} y_0(x)
\\
 y_1(x) 
\end{array}
\right)
\label{M2}
\end{equation}
which is in the form of Eq.(\ref{dydx}).
\cite{Khorasani}
Eigenvalues are determined by
\begin{equation}
 0 = \left\vert
\begin{array}{cc}
\lambda  & - 1 \\
 k^2(x) & \lambda
\end{array}\right\vert
= \lambda^2 + k^{2}(x)
\,,
\end{equation}
from which we get
\begin{equation}
 \lambda_1 = i k \,,\quad \lambda_2 = -i k
\,.
\label{WKB0}
\end{equation}
Eigenvectors and dual eigenvectors are 
\begin{equation}
 \ket{\lambda_1} = 
\left(
\begin{array}{c}
1 \\ i k
\end{array}\right)
\,,\quad
 \ket{\lambda_2} = 
\left(
\begin{array}{c}
1 \\ - i k
\end{array}\right)
\,,
\end{equation}
\begin{equation}
 \bra{\lambda_1} = \frac{1}{2} \left( 1 \,,\, - \frac{i}{k} \right)
\,,\quad
 \bra{\lambda_2} = \frac{1}{2} \left( 1 \,,\, \frac{i}{k} \right)
\,.
\end{equation}
Then we get
\begin{equation}
 \frac{dS_1}{dx} = - 
 \bra{\lambda_1}\frac{d}{dx}\ket{\lambda_1} = 
 - \frac{1}{2} \left( 1 \,,\, - \frac{i}{k} \right)
\left(
\begin{array}{c}
0 \\ i \frac{dk}{dx} 
\end{array}\right)
= - \frac{1}{2} \frac{d}{dx}\log k
\,.
\end{equation}
The approximate wave function up to the 1st order is
\begin{equation}
\psi(x) = y_0(x) \approx 
\exp\left( \frac{1}{\epsilon} S_{0} + S_1 \right) 
= \frac{1}{\sqrt{k(x)}} \exp\left(
i \int^{x} k(s) ds
\right)
\label{WKBwaveFunction}
\end{equation}
which is nothing but a usual WKB wave function.
The adiabatic approximation breaks down at the level crossing 
point that is equal to the classical turning point ($k(x) = 0$).
The breakdown manifests itself in 
the divergence of the approximate wave function, Eq.(\ref{WKBwaveFunction})
 at the classical turning point.

Various approaches have been devised in order to overcome 
this divergence problem.
\cite{cubicWKBmethod,cubicWKBtheory,Khorasani,FiniteWKB}
Especially, 
the authors of Ref.s ~\onlinecite{cubicWKBmethod,cubicWKBtheory} have 
successfully obtained wave functions without divergence 
in the whole coordinate range.
Hence they named this method ``divergence-free WKB''. 
The simplest version of this method is called ``cubic-WKB''
because the 0th-order wave functions are built with the use of 
roots of a certain cubic algebraic equation (see Eq.(\ref{cubicWKBeq})).
The outline of this method is presented below: 

Introducing $\varphi(x)$ through
\begin{equation}
\psi(x) = e^{\varphi(x)}
\,,
\end{equation}
Eq.~(\ref{SchroedingerEq}) reduces to 
\begin{equation}
 (\varphi')^2 + k^2 = - \varphi''
\label{phi2}
\,.
\end{equation}
When the term $\varphi''$ is neglected, 
Eq.(\ref{phi2}) gives the 0th-order WKB solutions, 
Eq.(\ref{WKB0}).

One more differentiation of Eq.(\ref{phi2}),
\begin{equation}
 2 \varphi' \varphi'' + 2 k k' = - \varphi'''
\,,
\end{equation}
is again inserted into Eq.(\ref{phi2}), 
and we get
\begin{equation}
 (\varphi')^3 + k^2  \varphi' - k k' = \frac{\varphi'''}{2}
\,.
\label{phi3}
\end{equation}
Regarding $\varphi'''/2$ as a higher order term, 
the following cubic equation
\begin{equation}
 (\varphi')^3 + k^2  \varphi' - k k' = 0
\label{cubicWKBeq}
\end{equation}
gives the 0th-order ``cubic-WKB'' solutions.
Combining three roots of Eq.(\ref{cubicWKBeq}), 
the authors of 
 Ref.s~\onlinecite{cubicWKBmethod,cubicWKBtheory} have 
succeeded in obtaining an approximate wave function without 
divergence in the whole range of $x$.

We now try to express 
the above result in the form of Eq.(\ref{ExpansionOfc1}).
Since the key equation (\ref{cubicWKBeq}) is cubic 
instead of quadratic, 
we need a $3\times 3$ matrix $M(x)$ and , in turn , 
need one more equation in addition to Eq.(\ref{M2}).
Differentiating Eq.(\ref{SchroedingerEq}), we get
\begin{equation}
 \psi''' = - 2 k k' \psi - k^2 \psi'
\,.
\label{dpd3}
\end{equation}
In addition to $y_0$ and $y_1$, we introduce $y_2 = \psi''$.
Then (\ref{dpd3}) becomes
\begin{equation}
 y_2'  = - 2 k k'y_0 - k^2 y_1
\,.
\label{dy2dx}
\end{equation}
Since $y_1'$ ( $= \psi''$) is equal to both  $-k^2 y_0$ and 
$y_2$, there is room for one parameter which we denote as $\alpha$:
\begin{equation}
y_1' = \alpha y_2  - ( 1- \alpha) k^2 y_0
\,.
\label{dy1dx}
\end{equation}
The parameter $\alpha$ needs not to be a constant but can depend on $x$.
From Eqs. (\ref{dy2dx}) and (\ref{dy1dx}) 
together with $y_0' = y_1$, we get 
\begin{equation}
 \frac{d}{dx}\left(
\begin{array}{c}
 y_0 \\ y_1 \\ y_2
\end{array}\right)
= \left(
\begin{array}{ccc}
0 & 1 & 0 \\
 (\alpha -1 )k^2 & 0 & \alpha \\
- 2 k k' & - k^2 & 0 
\end{array}\right)
\left(
\begin{array}{c}
 y_0 \\ y_1 \\ y_2
\end{array}\right)
\,.
\label{M3}
\end{equation}
It should be noted that the solution of Eq.(\ref{SchroedingerEq}) 
is also a solution of Eq.(\ref{M3}) , but 
the converse statement does not hold.
Hence additional conditions are necessary to select proper 
solutions of Eq.(\ref{M3}). 
From Eq.(\ref{M3}), we get
\begin{equation}
\frac{d}{dx}\left(\frac{y_0'' + k^2 y_0}{\alpha}\right)=0
\,.
\end{equation}
Therefore, we can see that 
the solution of Eq.(\ref{M3}) with the 
initial condition  
satisfying $y_2(0) + k^2(0) y_0(0) = 0$
becomes the solution of Eq.(\ref{SchroedingerEq}).

The adiabatic eigenvalues of (\ref{M3}) are determined by 
\begin{equation}
 \lambda^3 + k^2 \lambda + 2 \alpha k k' = 0
\end{equation}
which coincides with Eq.(\ref{cubicWKBeq}) if $\alpha = - 1/2$.
The reason of this choice will be explained 
below Eq.(\ref{dS1dx}).

The eigenvector and dual eigenvector with eigenvalue $\lambda$ are 
\begin{equation}
\ket{\lambda} = 
\left(\begin{array}{c}
1 \\ \lambda \\ \frac{\lambda^2 + (1-\alpha)k^2}{\alpha}
\end{array}\right)
=
\left(\begin{array}{c}
1 \\ \lambda \\ - k^2 - \frac{2 k k'}{\lambda}
\end{array}\right)
\,,
\end{equation}
\begin{equation}
\bra{\lambda} = N^{-1} 
\left( \lambda + \frac{\alpha k^2}{\lambda}\,,\, 1 \,,\, 
\frac{\alpha}{\lambda}\right)\,,\quad N = 3 \lambda + \frac{k^2}{\lambda}
\,.
\end{equation}
Then, from Eq.(\ref{S1}),  the 1st-order term becomes 
\begin{equation}
\frac{dS_1}{dx} =
 - \bra{\lambda}\frac{d}{dx}\ket{\lambda}
= - \frac{3 \lambda \lambda' + 2 (1-\alpha) k k'
- \frac{\alpha'}{\alpha}(\lambda^2 + k^2)
}{3\lambda^2 + k^2}
\,.
\label{dS1dx0}
\end{equation}
Far away from the classical turning point, 
the eigenvalues which are necessary to build approximate 
wave functions are 
$\lambda \approx \pm i k$ and
\begin{equation}
 \lambda^2 \approx - k^2\,,\quad 
\lambda \lambda' \approx - k k'
\,.
\end{equation}
In this case, 
\begin{equation}
\frac{dS_1}{dx} \approx
 -  \frac{(1 +  2 \alpha) k'}{2 k}
\label{dS1dx}
\,.
\end{equation}
We can see the choice $\alpha = - 1/2$ makes
the 1st-order correction vanish approximately.
Since the smallness of the 1st-order correction usually 
means small errors of the lowest-order result, 
this choice seems reasonable.

\section{Conclusion}

In this work we have shown  WKB as well as 
``cubic-WKB'' methods can be  expressed 
in the form of an adiabatic expansion.
As for the ``cubic-WKB'' method, 
the adiabatic expansion of a wave function contains a certain parameter.
When this parameter is chosen so as to make 
the 1st-order term vanish approximately,
the lowest-order adiabatic wave function becomes 
equal to the lowest-order ``cubic-WKB'' wave function.

Comments on further study are in order;
Since the parameter $\alpha$ is in general a function of $x$,
we may be able to choose $\alpha(x)$ so that 
the 1st-order term vanishes not approximately but exactly if 
we focus on a specific eigenvector.
To examine what happens in this case would be worth pursuing.
An $N$th-order WKB method, the extension of ``cubic-WKB'', 
has been presented in Ref.~\onlinecite{cubicWKBtheory}.
To extend the present result toward this direction is interesting.
The authors of Refs.~\onlinecite{cubicWKBmethod,cubicWKBtheory} have furher 
extended their idea in the context of 
the steepest descent method.
\cite{SteepestDescentMethod-1,SteepestDescentMethod-2}
To investigate a relation between the present results and 
theirs seems also interesting.


\begin{acknowledgments}
The author would like to thank T. Hyouguchi for his valuable 
comment on the choice of $\alpha$.
\end{acknowledgments}

{\bf An additonal note}

After this manuscript was published in 
{\it Mod. Phys. Lett. A} {\bf 34},
1950250 (2019), 
I have noticed papers \cite{Mostafazadeh2014-1, Mostafazadeh2014-2} 
discussing the relation between 
WKB and adiabatic approximations. 
I thank the author of Ref.s~\onlinecite{Mostafazadeh2014-1} and \onlinecite{Mostafazadeh2014-2} for informing me of this work.



\begin{thebibliography}{0}
\bibitem{BerryMount} M.V. Berry and  K.E. Mount, 
{\it Rep. Prog. Phys.} 
{\bf 35}, 315
(1972). 

\bibitem{Miller} W.H. Miller, 
{\it Adv. Chem. Phys.}
{\bf 25}, 69
(1974).

\bibitem{SupersymmetryWKB}
R. Dutt, A. Khare, and U. P. Sukhatme,
{\it Am. J. Phys.} {\bf 59} 
, 723
(1991).

\bibitem{SupersymmetryWKB2}
L. Salasnich and F. Sattin,
{\it Mod. Phys. Lett. B} {\bf 11},
801
(1997).

\bibitem{ExactWKB}
E. Delabaere,
{\it J. Math. Phy. (N.Y.)} {\bf 38} 
, 6126
(1997).

\bibitem{Resurgence}
I. Gahramanov and K. Tezgin,
{\it Int. J. Mod. Phys. A} {\bf 32}, 1750033 (2017).

\bibitem{cubicWKBmethod} T. Hyouguchi, A. Adachi and M. Ueda, 
{\it Phys. Rev. Lett.} {\bf 88} 
, 170404
(2002).

\bibitem{cubicWKBtheory} T. Hyouguchi, R. Seto, M. Ueda, and  
A. Adachi,
{\it Ann. Phys. (N.Y.)} {\bf 312}, 
177
(2004). 

\bibitem{Messiah}
A. Messiah, {\it Quantum Mechanics}
 (North-Holland, Amsterdam, 1962).

\bibitem{GeometricPhase}
A. Shapere, F. Wilczek (Ed.), 
{\it Geometric Phases in Physics}
(World Scientific, Singapore, 1989).

\bibitem{Holstein}
B. R. Holstein,
{\it Am. J. Phys.} {\bf 57} 
,1079
(1989).

\bibitem{Khorasani}
S. Khorasani, 
{\it Sci. Iran. D} {\bf 23}, 2928 (2016).

\bibitem{FiniteWKB}
U. Sukhatme and A. Pagnamenta,
{\it Am. J. Phys.} {\bf 59} 
, 944
(1991).

\bibitem{SteepestDescentMethod-1}
T. Hyouguchi, R. Seto, and  S. Adachi, 
{\it Prog. Theor. Phys.} {\bf 122} 
, 1311
(2009).

\bibitem{SteepestDescentMethod-2}
T. Hyouguchi, R. Seto, and S. Adachi,
{\it  Prog. Theor. Phys.} {\bf 122} 
, 1347
(2009).

\bibitem{Mostafazadeh2014-1}
M. Mostafazadeh,
{\it J. Phys. A} {\bf 47}
, 125301
(2014).

\bibitem{Mostafazadeh2014-2}
M. Mostafazadeh,
{\it J. Phys. A} {\bf 47}
, 345302
(2014).
\end{thebibliography}
\end{document}